%% file: XRS_final_DM.tex
\documentclass[aps,prb,twocolumn,superscriptaddress]{revtex4-1}

\usepackage[utf8]{inputenc} 
\usepackage[english]{babel} 
\usepackage{graphicx} 
\usepackage{amssymb,amsmath} 
\usepackage[hidelinks]{hyperref} 
\usepackage{color}
\hypersetup{
	colorlinks,
	citecolor=blue,
	linkcolor=black,
	urlcolor=blue,
}

\begin{document}

\title{Crystal field splitting in Sr$_{n+1}$Ir$_n$O$_{3n+1}$ ($n=1,2$) iridates
\\ probed by x-ray Raman spectroscopy}

\author{M. Moretti~Sala}
\email{marco.moretti@esrf.fr}
\affiliation{European Synchrotron Radiation Facility, CS 40220, 38043 Grenoble
Cedex 9, France}

\author{M. Rossi}
\affiliation{European Synchrotron Radiation Facility, CS 40220, 38043 Grenoble
Cedex 9, France}

\author{A. Al-Zein}
\affiliation{European Synchrotron Radiation Facility, CS 40220, 38043 Grenoble
Cedex 9, France}

\author{S. Boseggia}
\affiliation{London Centre for Nanotechnology and Department of Physics and
Astronomy, University College London, London WC1E 6BT, United Kingdom}
\affiliation{Diamond Light Source Ltd, Diamond House, Harwell Science and
Innovation Campus, Didcot, Oxfordshire OX11 0DE, United Kingdom}

\author{E.~C. Hunter}
\affiliation{School of Physics and Astronomy, The University of Edinburgh,
Mayfield Road, Edinburgh EH9 2TT, United Kingdom}

\author{R.~S. Perry}
\affiliation{London Centre for Nanotechnology and Department of Physics and
Astronomy, University College London, London WC1E 6BT, United Kingdom}

\author{D. Prabhakaran}
\affiliation{Clarendon Laboratory, Department of Physics, University of Oxford,
Parks Road, Oxford OX1 3PU, United Kingdom}

\author{A.~T. Boothroyd}
\affiliation{Clarendon Laboratory, Department of Physics, University of Oxford,
Parks Road, Oxford OX1 3PU, United Kingdom}

\author{N.~B. Brookes}
\affiliation{European Synchrotron Radiation Facility, CS 40220, 38043 Grenoble
Cedex 9, France}

\author{D.~F. McMorrow}
\affiliation{London Centre for Nanotechnology and Department of Physics and
Astronomy, University College London, London WC1E 6BT, United Kingdom}

\author{G. Monaco}
\affiliation{Dipartimento di Fisica, Universit\`a di Trento, via Sommarive 14,
38123 Povo (TN), Italy}

\author{M. Krisch}
\affiliation{European Synchrotron Radiation Facility, CS 40220, 38043 Grenoble
Cedex 9, France}

\begin{abstract}
Non-resonant Raman spectroscopy in the hard X-ray regime has been used to
explore the electronic structure of the first two members of the
Ruddlesden-Popper series Sr$_{n+1}$Ir$_{n}$O$_{3n+1}$ of iridates. By tuning the
photon energy transfer around 530 eV we have been able to explore the oxygen K
near edge structure with bulk sensitivity. The angular dependence of the spectra
has been exploited to assign features in the 528-535 eV energy range to specific
transitions involving the Ir $5d$ orbitals. This has allowed us to extract
reliable values for both the $t_{2g}$--$e_g$ splitting arising from the cubic
component of the crystal field ($10Dq$), in addition to the splitting of the
$e_g$ orbitals due to tetragonal distortions. The values we obtain are (3.8,
1.6) eV and (3.55, 1.9) eV for Sr$_2$IrO$_4$ and Sr$_3$Ir$_2$O$_7$,
respectively.
\end{abstract}

\maketitle

\section{Introduction}

$5d$ transition metal oxides have recently attracted considerable interest as
they display unusual properties primarily resulting from the effect of large
spin-orbit coupling
\cite{Kim2008,Moon2008,Kim2009,Pesin2010,Wang2011,KimJW2012,Watanabe2013,
Witczak2014}. Of particular interest is the electronic nature of
Sr$_2$IrO$_4$\cite{Crawford1994} and Sr$_3$Ir$_2$O$_7$\cite{Cao2002}: despite
the large $5d$ bandwidth and weak correlation, both of which favour a metallic
character, these systems are insulators. The opening of an electronic gap has
been explained by means of a Hubbard-like model, in which the effect of
correlation is enhanced by the strong spin-orbit coupling which narrows the
effective $5d$ bandwidth isolating the so-called $j_\mathrm{eff}=1/2$
state\cite{Kim2008,Kim2009}. The $j_\mathrm{eff}=1/2$ state results from a
particular hierarchy of energies at play, most especially the crystal field and
the spin-orbit coupling. 

Sr$_2$IrO$_4$ (Sr$_3$Ir$_2$O$_7$) is the $n=1$ ($n=2$) member of the
Ruddlesden-Popper series, Sr$_{n+1}$Ir$_{n}$O$_{3n+1}$, and is built by the
stacking of IrO$_2$ (bi-)layers, in which IrO$_6$ octahedra share the corner
oxygens. The dominant perturbation to the half-filled $5d$ iridium states 
in these compounds comes from the cubic component of the crystal field, written
conventionally as $10Dq$. Indeed the $t_{2g}$--$e_g$ splitting, of order 
several eV, is often considered to be large enough that the $e_g$ states can be
neglected, allowing the basic electronic structure to be understood in terms of
a single hole occupying the $t_{2g}$ orbitals (tetravalent iridium is $5d^5$).
In order to describe properly the ground state wave function of this hole,
spin-orbit coupling and residual crystal-field effects with symmetry lower than
cubic, such as tetragonal in the post-perovskite
CaIrO$_3$\cite{Hirai2009,Ohgushi2013} or trigonal in pyrochlore R$_2$Ir$_2$O$_7$
(R = rare earth element)\cite{Hozoi2012}, need to be considered. At the
single-ion level, this is achieved by diagonalizing the Hamiltonian $\mathcal{H}
= \zeta \mathbf{L}\cdot\mathbf{S}-\Delta  L_z^2$ in the
$t_{2g}$ orbitals
basis-set\cite{Ament2011,Liu2012,Hozoi2012,Ohgushi2013,Moretti2014PRL}, where
$\zeta$ is the spin-orbit coupling and $\Delta$ is the tetragonal (trigonal)
crystal-field splitting. 

Strictly speaking, the $j_\mathrm{eff}=1/2$ ground state is realized only for
$\Delta=0$, i.e. for a perfectly cubic symmetry. In real materials this
condition is relaxed to $|\Delta|\ll\zeta$. Estimates of $\Delta$ in
Sr$_2$IrO$_4$ ($\Delta=-0.01$ eV\cite{Boseggia2013JPCM}) and its sister
compound Ba$_2$IrO$_4$($\Delta=0.05$ eV\cite{Moretti2014PRB}) indeed confirm
that the requirement on the relative magnitude of $|\Delta|$ and $\zeta$ is
realized, since the spin-orbit coupling in these materials  of order
$\sim$0.5 eV\cite{Kim2008,Boseggia2013PRL,Moretti2014PRB}. One has to keep in
mind, however, that the scenario of the $j_\mathrm{eff}=1/2$ ground state holds
true only when the $e_g$ states do not contribute to the ground state wave
function, i.e. if the cubic component of the crystal field $10Dq$ is much larger
than the spin-orbit coupling, $10Dq\gg\zeta$. Indeed, the contribution of the
$e_g$ states has been invoked as a possible cause of the departure of CaIrO$_3$
from the pure $j_\mathrm{eff}=1/2$ ground state in LDA+SO+U
calculations\cite{Subedi2012}.

Theoretical estimates of $10Dq$ in Sr$_2$IrO$_4$ range from 1.8\cite{Haskel2012}
to 5 eV\cite{Jin2009}. Experimentally, various x-ray techniques have been used
to estimate $10Dq$, including x-ray absorption spectroscopy (XAS),
resonant elastic (REXS) and inelastic (RIXS) x-ray scattering. For example, soft
XAS at the O K edge has been used to probe the empty iridium 5$d$ states through
hybridization with the oxygen 2$p$ orbitals\cite{Kim2008,Moretti2014PRB},
providing values of $10Dq$ for Sr$_2$IrO$_4$\cite{Moon2006} and
Sr$_3$Ir$_2$O$_7$\cite{Park2014} in the range 2.5 eV to 4 eV. However, this
particular technique is highly surface sensitive, especially when performed in
total-electron-yield (TEY) mode, which compromises the reliability of the
extracted value of $10Dq$. The possibility that surface and bulk properties
might be different in iridium oxides was highlighted by Liu \textit{et al.},
who reported the existence of weak metallicity in the
near-surface electronic structure of Sr$_3$Ir$_2$O$_7$ while its bulk is known to be
insulating\cite{Liu2014}. In addition to the surface sensitivity, one has to
deal with self-absorption effects in total-fluorescence-yield (TFY) detected
XAS. As self-absorption is dependent on photon energy and experimental geometry,
extreme caution has to be taken when corrections to the spectra are applied. XAS
at the Ir L$_{2,3}$ edges ensures bulk-sensitivity, but self-absorption equally
affects hard XAS in TFY mode. Moreover, it suffers from the sizeable broadening
of features due to the 2$p$ core-hole lifetime which obscures
details of the electronic structure close to the Fermi energy. This problem can
at least be overcome to a certain degree by measuring partial-fluorescence-yield (PFY)
detected XAS\cite{Hamalainen1991}: this technique provides very
similar information to that of conventional XAS, but with the advantage that a
shallower core-hole is left in the final state of the decay process selected by
energy-discriminating the photons emitted due to radiative decay. For
example, in the case of the $L\alpha_{1}$ ($L\alpha_{2}$) emission line of iridium,
if $\Gamma_{2p}$ is the lifetime broadening of the $2p_{3/2}$ core-hole, and
$\Gamma_{3d}$ is that of the $3d_{5/2}$ ($3d_{3/2}$) core-hole, then the PFY
broadening will be given by
$1/\sqrt{1/\Gamma_{2p}^{2}+1/\Gamma_{3d}^{2}}\approx\Gamma_{3d}$, since
$\Gamma_{3d}\ll\Gamma_{2p}$. However, even if the benefits of PFY XAS are
evident, it is still difficult to extract quantitative information on $10Dq$
from such measurements\cite{Gretarsson2011,Clancy2014}. 

Resonant x-ray magnetic scattering
(RXMS)\cite{Kim2009,KimJW2012,Boseggia2012JPCM,Boseggia2013JPCM} and
resonant inelastic x-ray scattering
(RIXS)\cite{Ishii2011,KimJ2012,Liu2012,Moretti2014PRL2} in the hard x-ray regime
also provide rough estimates of the cubic component of the crystal field from
the RXMS/RIXS energy dependence. Indeed, that the intensity of both magnetic
reflections in RXMS and intra-$t_{2g}$ excitations in RIXS are enhanced a few eV
below the main absorption line has been interpreted as a signature of the
$t_{2g}$--$e_g$ splitting. Again, however, both of these techniques suffer from
self-absorption effects due to the proximity of the scattered photon energy to
the Ir L$_{2,3}$ absorption edges. 

The present work was designed to provide a reliable, bulk-sensitive probe of the
electronic structure of iridium oxides. We therefore used non-resonant inelastic
x-ray scattering (NIXS) in the hard x-ray energy range, more specifically x-ray
Raman spectroscopy (XRS), to probe the bulk properties of iridium oxides. XRS is
a x-ray scattering technique in which the energy of the incoming and scattered
photons is far from absorption edges of the material, making XRS a
self-absorption-free and bulk-sensitive probe\cite{Schulke2007}. Indeed, the XRS
cross-section in the limit of small momentum transfer $|\mathbf{q}|$ (i.e. in
the dipole limit) is formally identical to that of XAS, with $\mathbf{q}$
playing the role of photon polarization: the XRS cross-sections is then
proportional to $\left| \langle f | \mathbf{q} \cdot \mathbf{r} | i \rangle
\right|^2$, where $|i\rangle$ and $|f\rangle$ are the many-body electronic wave
functions of the initial and final state of the system,
respectively\cite{Schulke2007}. The main drawback of this technique is the low
count-rate, which is partially overcome by collecting the scattered photons over
a large solid angle. In the following we show that XRS allows the precise
determination of the cubic component of the crystal-field splitting in the
compounds Sr$_2$IrO$_4$ and Sr$_3$Ir$_2$O$_7$, thus offering an alternative
spectroscopic tool for the investigation of the electronic structure of iridium
oxides. 

\section{Experimental details}

X-ray Raman spectroscopy measurements were performed at the ID20 beam line of
the European Synchrotron Radiation Facility (ESRF), Grenoble. The X-rays
produced by four U26 undulators were monochromatized to an energy-resolution
of $\Delta E_\mathrm{i} \simeq 0.3$ eV by the simultaneous use of a
Si(111) high heat-load liquid-nitrogen cooled monochromator and a Si(311)
post-monochromator. The x-rays were then focused at the sample position by
means of a Kirkpatrick-Baez mirror system down to a spot size of $10 \times 20$
$\mu$m$^2$ (vertical $\times$ horizontal, FWHM). The scattered X-rays were
collected by 12 crystal-analyzers exploiting the Si(660) reflection close to
backscattering geometry (at a fixed Bragg angle of $88.5^\circ$, corresponding
to $E_\mathrm{o}=9670$ eV) and detected by a Maxipix detector\cite{Ponchut2011}
with pixel size of $55\times 55$ $\mu$m$^2$. The resulting energy resolution was
$\Delta E \simeq 0.7$ eV. In order to obtain the XRS spectrum, the incident
photon energy $E_\mathrm{i}$ was varied in the energy range from
$E_\mathrm{i}-E_\mathrm{o}=0$ (the elastic energy) to
$E_\mathrm{i}-E_\mathrm{o}=570$ eV, thus covering the oxygen K edge. The
accumulation time/spectrum was about 2 hours and several spectra were recorded
to improve the counting statistics. XRS spectra were collected in two different
scattering geometries, corresponding to the momentum transfer $\mathbf{q}$ along
the sample $c$-axis and in the $ab$-plane, respectively. In both geometries, the
scattering plane was vertical and the incident X-rays linearly polarized in the
horizontal plane. XAS spectra were recorded at the ID08 beam line of the ESRF in
the TFY mode. 

Single crystals of Sr$_2$IrO$_4$ and Sr$_3$Ir$_2$O$_7$,  with
dimensions of $\sim 0.5 \times 0.5 \times 0.2$ mm$^3$, were 
were grown using
the flux method described in Ref. \onlinecite{Boseggia2012PRB}. 
All spectra were recorded at room temperature.

\section{Results and discussion}

\begin{figure}[t]
\centering
\includegraphics[width=.99\columnwidth]{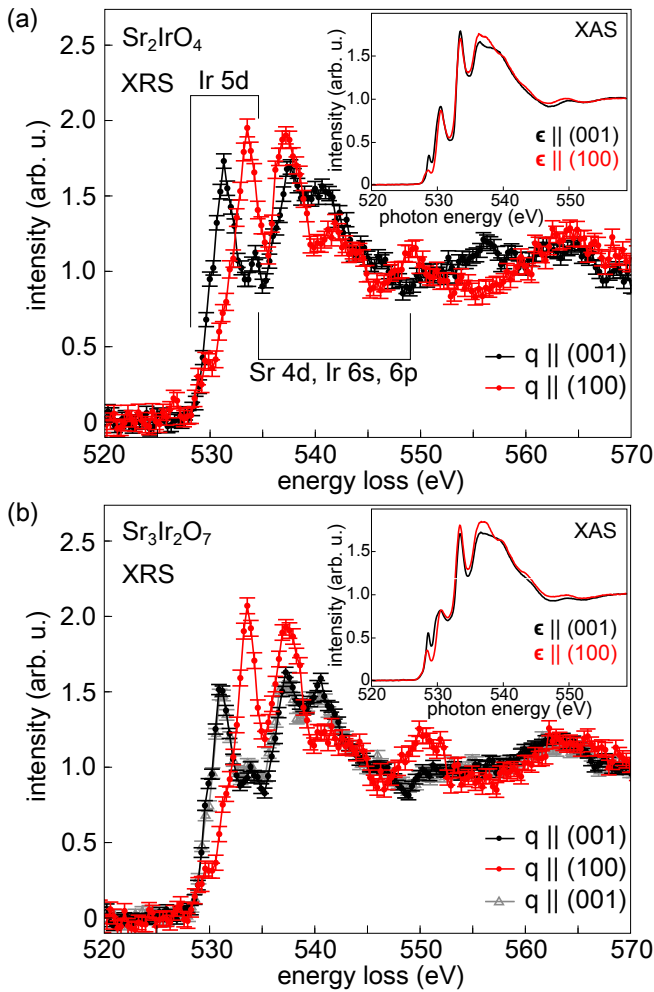}
\caption{\label{fig:fig1}XRS spectra of (a) Sr$_2$IrO$_4$ and (b)
Sr$_3$Ir$_2$O$_7$ for transferred momenta $\mathbf{q}\parallel(001)$ (black) and
$\mathbf{q}\parallel(100)$ (red dots) with $|\mathbf{q}|\simeq 6$ \AA$^{-1}$
(scattering angle $2\theta = 60^\circ$). Gray triangles in (b) represent the XRS
spectrum of Sr$_3$Ir$_2$O$_7$ with $|\mathbf{q}|\simeq 10$ \AA$^{-1}$
(scattering angle $2\theta = 120^\circ$). XAS spectra at the O K edge of the two
compounds for incoming polarization $\boldsymbol{\epsilon}\parallel(001)$
(black) and $\boldsymbol{\epsilon}\parallel(100)$ (red line) are also shown in
the insets.}
\end{figure}

Figure~\ref{fig:fig1} shows XRS scans for Sr$_2$IrO$_4$ (a) and
Sr$_3$Ir$_2$O$_7$ (b) across the oxygen K edge for $\mathbf{q} \parallel (001)$
(black) and $\mathbf{q}\parallel (100)$ (red dots). The scattering angle was
fixed to $2\theta=60^\circ$, corresponding to a momentum transfer of $|\mathbf{q}|\simeq
6$ \AA$^{-1}$. The background was removed by subtracting a linear fit to the
pre-edge region at energies lower than 528 eV. The spectra were then normalized
to unit area. For both samples, spectra taken in the two geometries are
distinctly different, revealing a very strong orientation dependence of the XRS
signal. In particular, one notes a large change of spectral weight between the
two main features in the 530-535 eV energy range. 

In agreement with XAS results\cite{Chen1991,Schmidt1996,Moon2006}, the 528-535
eV energy region is dominated by transitions to the Ir 5$d$ states through the
hybridization with O 2$p$ orbitals, while higher energy features correspond to
excitations involving Ir 6$s$, 6$p$ and Sr 4$d$ states\cite{Mizokawa2001}, as
indicated in Fig.~\ref{fig:fig1}. For comparison, TFY XAS spectra were measured
on the very same samples. These are shown in the insets of Fig.~\ref{fig:fig1}.
Continuous black and red lines correspond to orthogonal directions of the
photon polarization, $\boldsymbol{\epsilon}\parallel(001)$ and
$\boldsymbol{\epsilon}\parallel(100)$, respectively. As expected, the overall
shape is similar to that of the XRS spectra, but the dichroic effect in the XAS
spectra is very small, in stark contrast to the strong orientation dependence of
the XRS measurements performed on the same samples. In order to rule out any
contribution higher than dipolar to the XRS spectra, we investigated the
$|\mathbf{q}|$ dependence of the XRS cross-section in Sr$_3$Ir$_2$O$_7$: by
setting $2\theta=120^\circ$, corresponding to $|\mathbf{q}|\simeq 10$ \AA$^{-1}$
(gray triangles in Fig.~\ref{fig:fig1}(b)), we note that the overall shape of
the spectrum perfectly matches with that acquired for $|\mathbf{q}|\simeq 6$
\AA$^{-1}$, thus implying that the momentum dependence of the XRS is negligible.
We therefore attribute the discrepancy between XRS and XAS measurements to
potential surface and/or self-absorption effects affecting soft x-ray
techniques. This observation underlines the importance of complementing
surface-sensitive techniques with bulk-sensitive probes. 

In order to analyse our data we have calculated the number of 
peaks expected in the 530-535 eV energy interval and their corresponding spectral weights  by
pursuing the analogy between the XRS and XAS cross-sections.
The relevant transitions are those from O $1s$ to $2p$ states with the 
latter hybridised with the Ir 5$d$ orbitals\cite{Kim2008,Park2014,Moretti2014PRB}. 
The hybridization strength is calculated according to
the orbital overlap model \cite{Slater1954} with the hopping integral $t_{pd\mu}$
written as
\begin{equation}
t_{pd\mu}  = V_{pd\mu} r^{-\alpha},
\end{equation}
where $V_{pd\mu}$ is a constant depending on the bond type ($\mu$=$\pi$ or $\sigma$), $r$ is the Ir-O
distance ($r_\mathrm{A} = 2.06$ \AA\ and $r_\mathrm{P} = 1.98$ \AA\ for
apical and in-plane oxygens, respectively, in Sr$_2$IrO$_4$\cite{Crawford1994}; while $r_\mathrm{A} =
2.02$ \AA~and $r_\mathrm{P} = 1.99$ \AA~in
Sr$_3$Ir$_2$O$_7$\cite{Subramanian1994}) and $\alpha = 3.5$\cite{Harrison1989}.
It should be noted that $V_{pd\sigma}$ and $V_{pd\pi}$ are related by
$V_{pd\pi}=-V_{pd\sigma}/\sqrt{3}$\cite{Harrison1989}. 

\begin{figure}[t]
\centering
\includegraphics[width=.99\columnwidth]{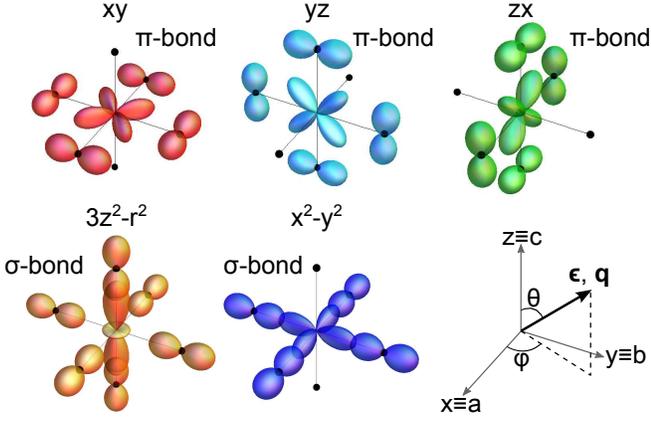}
\caption{\label{fig:fig2}Sketch of the symmetry of the $t_{2g}$ (top)
and $e_g$ (bottom) orbitals involved in the O $2p$-Ir $5d$ hybridization. The
Eulerian angles $\theta$ and $\varphi$ describing the direction of
$\boldsymbol{\epsilon}$  or $\mathbf{q}$ in the sample reference system are also
shown.}
\end{figure}

Since the hybridization strength is inversely proportional to the distance
between the atoms involved, we can distinguish the contributions of the
apical (A) and in-plane (P) oxygens. Let us consider the apical oxygens first:
the O $2p_\mathrm{z}$ state hybridizes with the Ir $5d$ $\mathrm{3z^2-r^2}$
states, while the $2p_\mathrm{x}$ ($2p_\mathrm{y}$) mixes with the $\mathrm{zx}$
($\mathrm{yz}$) orbitals. For the in-plane oxygens, 
$2p_\mathrm{z}$ hybridizes with the $\mathrm{yz}$ and $\mathrm{zx}$ orbitals,
while $2p_\mathrm{x}$ and $2p_\mathrm{y}$ are mixed with the $\mathrm{xy}$,
$\mathrm{3z^2-r^2}$ and $\mathrm{x^2-y^2}$ orbitals. This
is summarized in Fig.~\ref{fig:fig2}.
 
It remains to consider the cross-sections associated with  transitions to
different orbitals.
In the framework of a single-ion model, these are obtained by
calculating the matrix elements corresponding to the dipolar
$1s\rightarrow 2p_i$ transitions ($i=\mathrm{x, y, z}$)\cite{Moretti2014PRB}.
The cross-section is proportional to 
the product of $\left| t_{pd\mu} \right|^2$, $n$ the number of available final $5d$ states and 
a polarisation factor. 
The polarization (transferred momentum) dependence of the XAS (XRS, in the
dipole limit) cross-sections to the $2p_\mathrm{x}$, $2p_\mathrm{y}$ and
$2p_\mathrm{z}$ states are given by $\sin^2\theta\cos^2\varphi$,
$\sin^2\theta\sin^2\varphi$ and $\cos^2\theta$, respectively, where $\theta$ and
$\varphi$ are the Eulerian angles describing the direction of
$\boldsymbol{\epsilon}$ ($\mathbf{q}$) in the sample reference system, as
sketched in Fig.~\ref{fig:fig2}. Merging the cross-section angular dependence
and the hybridization between Ir $5d$-O $2p$ states, we obtain the polarization
(transferred momentum) dependence of the transitions to the $\mathrm{xy}$,
$\mathrm{yz}$, $\mathrm{zx}$, $\mathrm{3z^2-r^2}$, $\mathrm{x^2-y^2}$ orbitals
as reported in Table \ref{tab:table1}. Note that we have used
$n_\mathrm{xy}=n_\mathrm{yz}=n_\mathrm{zx} = 1/3$ and
$n_\mathrm{3z^2-r^2}=n_\mathrm{x^2-y^2} = 2$ expected for  the $j_\mathrm{eff}=1/2$
state.

\begin{table}[t]
\caption{\label{tab:table1}Polarization dependence of the O $1s
\rightarrow$ O $2p$-Ir $5d$ dipolar transitions.}
\begin{ruledtabular}
\begin{tabular}{c c c}
& Apical O & In-plane O \\
\hline
$\mathrm{xy}$ & 0 & $2V_{pd\pi}^2 
n_\mathrm{xy}r_\mathrm{P}^{-2\alpha}\sin^2\theta$
\\
$\mathrm{yz}$ &
$2V_{pd\pi}^2 n_\mathrm{yz}r_\mathrm{A}^{-2\alpha}\sin^2\theta\sin^2\varphi$ &
$2V_{pd\pi}^2 n_\mathrm{yz}r_\mathrm{P}^{-2\alpha}\cos^2\theta$ \\
$\mathrm{zx}$ &
$2V_{pd\pi}^2 n_\mathrm{yz}r_\mathrm{A}^{-2\alpha}\sin^2\theta\cos^2\varphi$ &
$2V_{pd\pi}^2 n_\mathrm{yz}r_\mathrm{P}^{-2\alpha}\cos^2\theta$ \\
$\mathrm{3z^2-r^2}$ &
$2V_{pd\sigma}^2 n_\mathrm{3z^2-r^2}r_\mathrm{A}^{-2\alpha}\cos^2\theta$ &
$V_{pd\sigma}^2 n_\mathrm{3z^2-r^2}r_\mathrm{P}^{-2\alpha}\sin^2\theta$ \\
$\mathrm{x^2-y^2}$ & 0 &
$\sqrt{3}V_{pd\sigma}^2 n_\mathrm{x^2-y^2}r_\mathrm{P}^{-2\alpha}\sin^2\theta$
\\
\end{tabular}
\end{ruledtabular}
\end{table}  

\begin{figure}[t]
\centering
\includegraphics[width=.99\columnwidth]{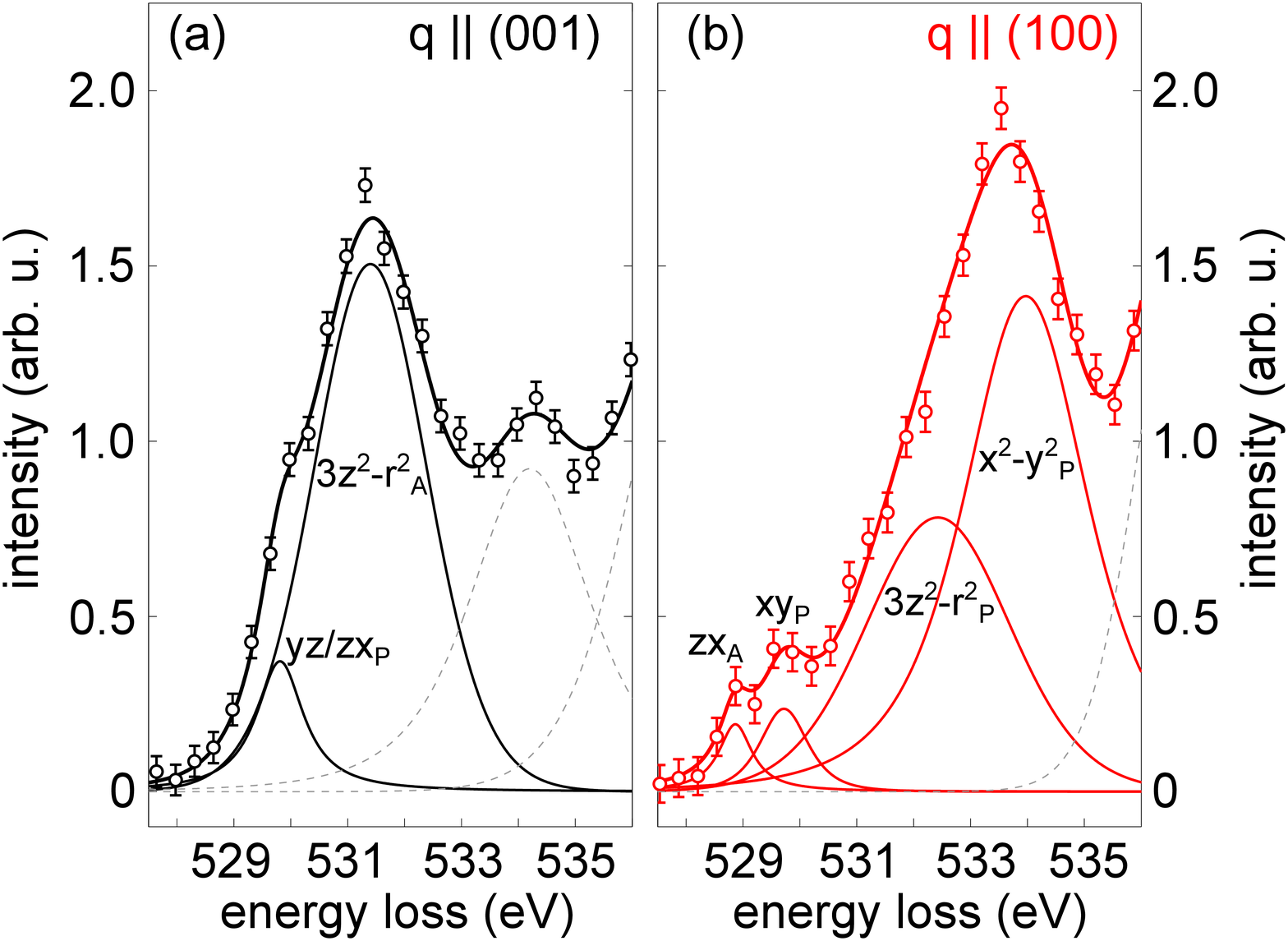}
\caption{\label{fig:fig3}Experimental (open dots) and constrained fit to
the XRS spectra (solid thick line) of Sr$_2$IrO$_4$ for (a) $\mathbf{q}\parallel
(001)$ and (b) $\mathbf{q}\parallel (100)$. The fitting curves are plotted in
solid lines, while the extra-peaks are reported in dashed gray lines.}
\end{figure}

\begin{figure}[t]
\centering
\includegraphics[width=.99\columnwidth]{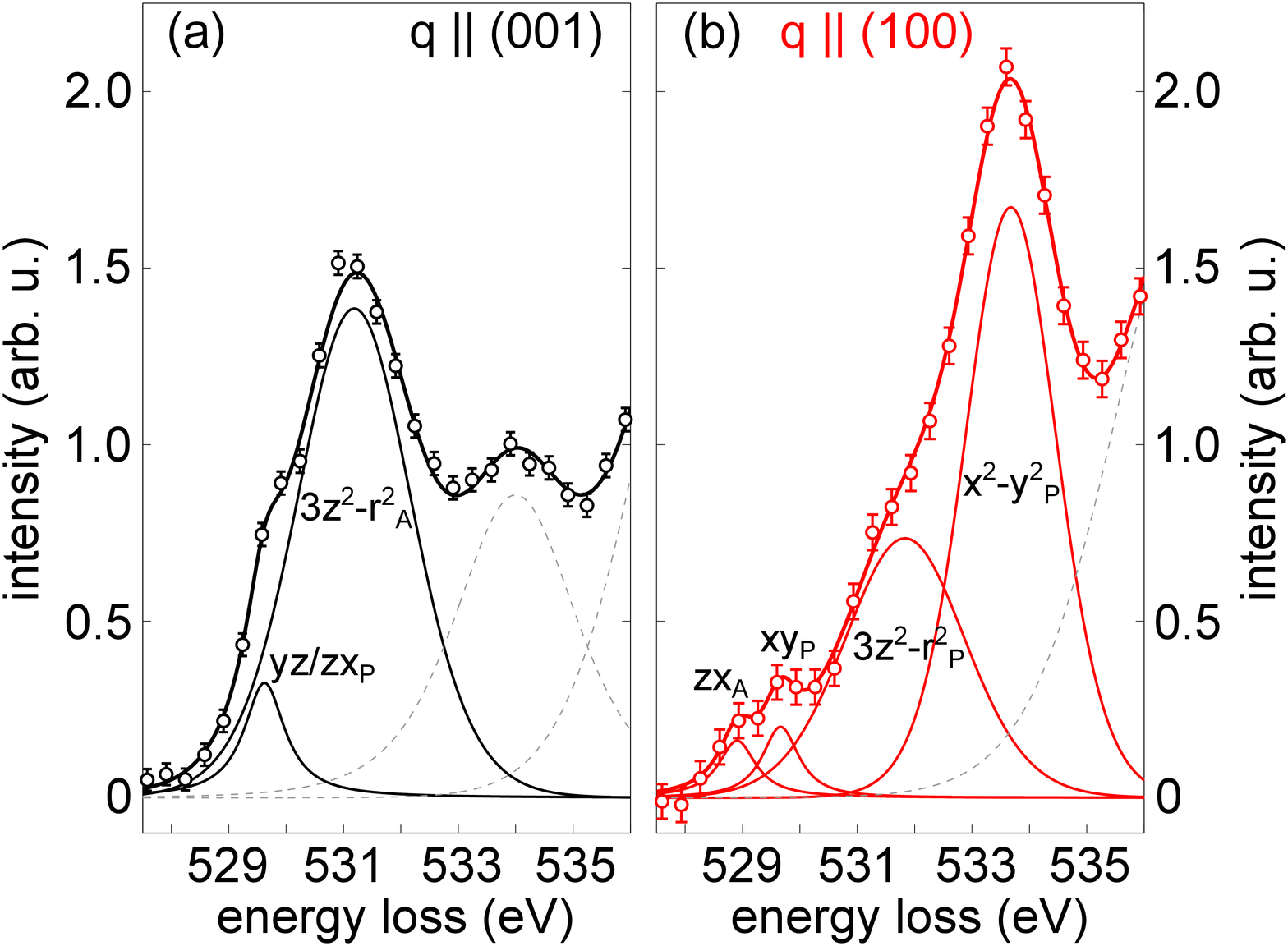}
\caption{\label{fig:fig4}Experimental (open dots) and constrained fit to
the XRS spectra (solid thick line) of Sr$_3$Ir$_2$O$_7$ for (a)
$\mathbf{q}\parallel (001)$ and (b) $\mathbf{q}\parallel (100)$. The fitting
curves are plotted in solid lines, while the extra-peaks are reported in dashed
gray lines.}
\end{figure}

For the specific geometries used in our experiments, 
it transpires that only two transitions are allowed when
$\mathbf{q}\parallel (001)$ ($\theta = 0$) and four when $\mathbf{q}\parallel
(100)$ ($\theta = 90^{\circ}$ and $\varphi = 0$). The appropriate cross-sections
are given in Table \ref{tab:table2}. We therefore performed a fitting  of our model
to the data
by adjusting the number of peaks accordingly and constraining their
relative spectral weight to the calculated one. Extra peaks were
introduced in the fit to mimic the high energy features: one for $\mathbf{q}\parallel
(100)$ and two for $\mathbf{q}\parallel (001)$, respectively. The result of the fitting is
shown in Fig.~\ref{fig:fig3} for Sr$_2$IrO$_4$ and in Fig.~\ref{fig:fig4} for
Sr$_3$Ir$_2$O$_7$. We emphasise 
that, apart from an overall scale factor for the amplitude, the energy
position and full width at half maximum (FWHM) of the curves are the only free
fitting parameters: their values are summarized in Table~\ref{tab:table2}. The
agreement between the fit and the experimental data is remarkably good in both
scattering geometries, allowing us to unambiguously assign each feature. 
In particular, the intense features at 531.4 (531.2) and 534.0 (533.7) eV in
Sr$_2$IrO$_4$ (Sr$_3$Ir$_2$O$_7$) correspond to excitations to the $\mathrm{3z^2-r^2}$ and
$\mathrm{x^2-y^2}$ orbitals via the apical and in-plane oxygens, respectively. This peak
assignment is consistent with the work of Moon \textit{et al.} on
Sr$_2$IrO$_4$\cite{Moon2006}, Schmidt \textit{et al.} on
Sr$_2$RuO$_4$\cite{Schmidt1996} and Park \textit{et al.} on
Sr$_3$Ir$_2$O$_7$\cite{Park2014}.


\begin{table*}
\caption{\label{tab:table2}Cross-sections, fitted energy positions and FWHM of
electronic dipolar transitions in Sr$_2$IrO$_4$ and Sr$_3$Ir$_2$O$_7$.}
\begin{ruledtabular}
\begin{tabular}{c c c c c c c}
& $\mathbf{q}\parallel (001)$ & $\mathbf{q}\parallel (100)$ & Energy loss (eV) &
FWHM (eV) & Energy loss (eV) & FWHM (eV) \\
& & & Sr$_2$IrO$_4$ & Sr$_2$IrO$_4$ & Sr$_3$Ir$_2$O$_7$ & Sr$_3$Ir$_2$O$_7$ \\
\hline
$\mathrm{xy/yz/zx_A}$ & & $2V_{pd\pi}^2 n_\mathrm{yz}r_\mathrm{A}^{-2\alpha}$ &
$528.9\pm 0.11$ & $0.71\pm 0.35$ & $528.9\pm 0.10$ & $0.78\pm 0.30$ \\
$\mathrm{xy/yz/zx_P}$ & $4V_{pd\pi}^2 n_\mathrm{yz}r_\mathrm{P}^{-2\alpha}$ &
$2V_{pd\pi}^2 n_\mathrm{xy}r_\mathrm{P}^{-2\alpha}$ & $529.8\pm 0.05$ & $1.0\pm
0.17$ & $529.6\pm 0.03$ & $0.87\pm 0.10$ \\
$\mathrm{3z^2-r^2_A}$ & $2V_{pd\sigma}^2
n_\mathrm{3z^2-r^2}r_\mathrm{A}^{-2\alpha}$
& & $531.4\pm 0.05$ & $2.4\pm 0.18$ & $531.2\pm 0.05$ & $2.4\pm 0.15$\\
$\mathrm{3z^2-r^2_P}$ & &
$V_{pd\sigma}^2 n_\mathrm{3z^2-r^2}r_\mathrm{P}^{-2\alpha}$ & $532.4\pm 0.75$ &
$3.0\pm 0.67$ & $531.8\pm 0.12$ & $2.5\pm 0.26$\\
$\mathrm{x^2-y^2_A}$ & & & & & & \\
$\mathrm{x^2-y^2_P}$ & &
$\sqrt{3}V_{pd\sigma}^2 n_\mathrm{x^2-y^2}r_\mathrm{P}^{-2\alpha}$ & $534.0\pm
0.35$ & $2.6\pm 0.46$ & $533.7\pm 0.05$ & $1.9\pm 0.06$\\
\end{tabular} 
\end{ruledtabular}
\end{table*} 

We are now in a position to extract the cubic component of the crystal field
$10Dq$. This is given by the energy difference between the centres of mass of
the $e_g$ and $t_{2g}$ states for in-plane oxygens. In view of the small
tetragonal crystal field measured in Sr$_2$IrO$_4$ ($|\Delta|=0.01$
eV\cite{Boseggia2013JPCM}), we consider the splitting of the $t_{2g}$ states due
to spin-orbit coupling only in the calculation of $10Dq$. We obtain $3.80\pm
0.82$ eV in Sr$_2$IrO$_4$ and $3.55\pm 0.13$ eV in Sr$_3$Ir$_2$O$_7$, assuming
$\zeta\simeq 0.4$ eV\cite{Kim2008}. Estimates of $10Dq$ extracted from XAS and
RXMS/RIXS measurements are consistent with our results. The cubic component of
the crystal field is thus very large compared to the other energy scales of the
system, namely the spin orbit coupling and the tetragonal crystal field,
therefore validating the initial hypothesis that $10Dq$ is the dominant energy
scale. Finally, in addition to the estimate of the cubic component of the
crystal field, we can deduce the sign of the tetragonal contribution to
the crystal field from the splitting of the $e_g$ states ($1.6\pm 0.82$ eV in
Sr$_2$IrO$_4$ and by $1.9\pm 0.13$ eV in Sr$_3$Ir$_2$O$_7$). Indeed, the fact
that the $\mathrm{x^2-y^2}$ orbital is the highest in energy is consistent with
structural studies indicating an elongation of the IrO$_6$ cage in both
compounds. Note that, for tetragonally distorted octahedra, the description of
$d$ states requires two parameters, $Ds$ and $D t$, in addition to the main
crystal-field parameter $10Dq$. The splitting of $e_g$ and $t_{2g}$ states is
then given by $4Ds+5Dt$ and $3Ds-5Dt$ ($=\Delta$),
respectively\cite{Bersuker2010}. In the absence of spin-orbit coupling,
the $t_{2g}$ states are almost degenerate ($\Delta \approx 0$), implying $3Ds
\approx 5Dt$. A finite splitting of the $e_g$ states is therefore compatible
with the realization of the $j_\mathrm{eff}=1/2$ ground state in Sr$_2$IrO$_4$
and Sr$_3$Ir$_2$O$_7$.

\section{Conclusions}  

By exploiting the orientation dependence of oxygen K edge XRS cross-sections in
Sr$_2$IrO$_4$ and Sr$_3$Ir$_2$O$_7$, we have been able to assign spectral
features in the 528-535 eV energy range to specific transitions involving the Ir
$5d$ orbitals. These assignments allow us to extract the value of the cubic
crystal-field splitting $10Dq$ of $3.80\pm 0.82$ and $3.55\pm 0.13$ eV in
Sr$_2$IrO$_4$ and Sr$_3$Ir$_2$O$_7$, respectively. In addition, the tetragonal
crystal field was found to split the $e_g$ states by $1.6\pm 0.82$ eV in
Sr$_2$IrO$_4$ and by $1.9\pm 0.13$ eV in Sr$_3$Ir$_2$O$_7$. It is important to
stress that the reliability of these values of the crystal field splittings
obtained in our study is enhanced by the bulk sensitivity of the XRS technique.

\section{Acknowledgments}  

The authors are grateful for technical support by C. Henriquet and R. Verbeni,
and all the colleagues from the ESRF support groups.

\input{XRS_final_DM.bbl}

\end{document}

%% file: XRS_final_DM.bbl
%